\documentclass[aps,prl,superscriptaddress,twocolumn]{revtex4-2}
\usepackage{graphicx}
\usepackage{dcolumn}
\usepackage{bm}
\usepackage{xcolor}

\begin{document}

\title{Searching for flavor dependence in nuclear quark behavior}

\author{J. Arrington}
\email{johna@anl.gov}
\affiliation{Physics Division, Argonne National Laboratory, Argonne, IL 60439}

\author{N. Fomin}
\email{nfomin@utk.edu}
\affiliation{University of Tennessee, Knoxville, TN 37996}

\date{\today}

\begin{abstract}

The observed correlation between the EMC effect and the contribution of short-range correlations (SRCs) in nuclei suggests that the modification of the quark distributions of bound protons and neutrons might occur within SRCs.
This raises the possibility that the EMC effect may have an isospin dependence arising from the $np$ dominance of SRCs.
We discuss previous attempts to test this possibility and perform a new analysis of existing data.
We find no experimental support for the observation of an isospin dependence of the EMC effect.
\end{abstract}

\keywords{Suggested keywords} 
\maketitle

The atomic nucleus is relatively well described as a collection of independent protons and neutrons, bound together through their mean-field interactions.
While this picture gives a remarkably good description of many properties of nuclei, there are a handful of observables that indicate a more complex underlying nature. In particular, measurements associated with energetic nucleons in nuclei, as well as studies of the nuclear quark distributions, show deviations from this simple but frequently effective approximation.

When the quark distributions in nuclei, represented by parton distribution functions (pdfs), were first measured~\cite{Aubert:1983xm}, it was expected that the nuclear pdfs would be nearly equal to the incoherent sum of the proton and neutron pdfs.
Deviations were expected to occur at large values of Bjorken-$x$, the fraction of the nucleon's longitudinal momentum carried by struck quark, due to smearing arising from the finite momentum of the bound nucleons. However, data from the EMC collaboration and later, other experiments~\cite{Aubert:1983xm, bodek83, gomez94} also observed a suppression of the quark distributions in the region $0.3 < x < 0.7$ in nuclei relative to the deuteron. This depletion of high-momentum quarks, known as the ``EMC effect'', increases with $x$ over this region, and grows with the mass of the nucleus.
Despite decades of study, the origin of the EMC effect is not yet known, but recent data showing identical nuclear dependence for high-momentum nucleons hint that the source of the EMC effect may live within these short-distance $NN$ configurations~\cite{Seely:2009gt, fomin2012, arrington11,Arrington:2015wja,Hen:2016kwk,Fomin:2017ydn}.

The short-distance part of the nucleon-nucleon interaction includes a hard short-range repulsive core as well as strong intermediate-range tensor attraction~\cite{Wiringa:1994wb}. These strong interactions between nucleons at short distance yield high-momentum components in the nucleon momentum distributions in nuclei that are not included in mean-field treatment~\cite{Benhar:1994hw, CiofidegliAtti:1995qe}. Inclusive electron scattering from nuclei at large momentum transfer and low energy transfer are sensitive to these high-momentum components~\cite{frankfurt81, frankfurt88, Sargsian:2002wc, arrington11}. The universal behavior of scattering from nuclei in these kinematics~\cite{frankfurt93, egiyan03, egiyan06, fomin2012} supports the hypothesis that these contributions come from two-body interactions that generate pairs of nucleons with high back-to-back momenta, referred to a short-range correlations (SRCs)~\cite{frankfurt88}. In addition, two-nucleon knockout measurements~\cite{shneor07, subedi08, Korover:2014dma} demonstrate that these high-momentum components are dominated by interactions between neutrons and protons, suggesting that the tensor interaction is the most important part of the $N$-$N$ interaction in these measurements~\cite{schiavilla06}.

Early observations of both the EMC effect~\cite{gomez94} and the presence of SRCs~\cite{frankfurt93, egiyan03, egiyan06} in nuclei showed an $A$ dependence that scaled roughly with the average nuclear density~\cite{frankfurt88, day89, sick92}. 
This situation changed with a new generation of measurements at Jefferson Laboratory, where precision measurements of both the EMC effect~\cite{Seely:2009gt} and strength of SRCs~\cite{fomin2012} were made in a set of light nuclei. First, it was observed that the EMC effect in $^9$Be was much larger than one would expect for this very low-density nucleus. It was pointed out that even if the EMC effect depends on density, the \textit{average} nuclear density may not be the most relevant quantity. $^9$Be has strong $\alpha$ clustering~\cite{Arai96}, with a significant component that looks like $^4$He nuclei plus an additional neutron. If one pictures $^9$Be as a 3-body system (two $\alpha$ clusters and one neutron), one obtains a low average density even though most of the nucleons are in the dense $\alpha$ clusters. Note that when discussing the size of the EMC effect, we will use the convention introduced in Ref.~\cite{Seely:2009gt}, the slope of $R_{EMC}(x)=(\sigma_A/A)/(\sigma_D/2)$ in the region where the ratio falls linearly, roughly $0.3<x<0.7$. This yields a measure of the relative drop of the large-$x$ structure function ratio relative to the `no EMC effect' value of $R_{EMC}=1$, under the assumption that the longitudinal-to-transverse cross section ratio is $A$ independent.

Shortly after this observation for the EMC effect, a similar pattern was observed in measurements of SRCs in light nuclei~\cite{fomin2012}.
This correlation between the number of SRC pairs and the size of the EMC effect suggested the possibility of a closer connection between these two effects. Broadly speaking, there are two classes of hypotheses that have been proposed~\cite{Arrington:2012ax}. The local density (LD) picture  assumes that the EMC effect is driven by the presence of nucleons in close proximity, e.g. through quark exchange in overlapping nucleons, and thus the EMC-SRC correlation exists because both phenomena are driven by short-distance configurations. The high virtuality (HV) picture supposes that the EMC effect is a result of high-momentum nucleons, e.g. due to large off-shell effects in highly-virtual nucleons, and is thus directly sensitive to the contribution of SRCs.

Understanding the connection between the EMC effect and SRCs can have significant implications. The dominance of $np$ pairs in SRCs suggests that the absolute contribution of high-momentum protons and neutrons in nuclei is nearly identical, meaning that the \textit{fractional} contributions can be different in non-isoscalar nuclei~\cite{Arrington:2015wja,Hen:2016kwk,Fomin:2017ydn}. If a larger fraction of the protons in a neutron-rich nucleus have high momenta, and the EMC effect is a result of these high-momentum nucleons (the HV picture), then the enhanced EMC effect for the protons will generate a flavor dependence, with $up$ quarks showing a greater modification than $down$ quarks. This has potential consequences for neutrino scattering and quark-level observables in non-isoscalar nuclei~\cite{cloet09, Cloet:2012td, cloet19}. Finally, an improved understanding of the nuclear effects in the deuteron is critical to the extraction of a wide range of neutron structure observables~\cite{Arrington:2011qt, Owens:2012bv}, and an improved understanding of the $A$ dependence of the EMC effect will allow us to constrain these effects in the deuteron.

The observed $np$ dominance of SRC pairs and its correlation to the size of the EMC effect is not the only reason to expect isospin dependence in the latter phenomenon. The expecation for the EMC effect in protons to be larger than for neutrons in neutron-rich nuclei has been predicted by calculations~\cite{cloet09,Cloet:2012td} and is a natural consequence of many simple pictures of the EMC effect~\cite{Arrington:2015wja, malace14, cloet19}. However, at this point there is no data that provides direct evidence for an isospin dependence in the EMC effect. While data show $np$ dominance for the high-momentum SRCs in nuclei, no data exist to tell us whether this
$np$ dominance also exists in the short-distance but low-momentum components of the nuclear wave function.
For this analysis, the EMC effect is assumed to be isospin independent in the LD picture, as the short-distance contributions are taken to be identical for $pp$, $np$, and $nn$ pairs. 

A previous work examined the quality of the observed proportionality between the SRCs and the EMC effect for the HV and LD models~\cite{Arrington:2012ax}. In the HV model, one assumes that only $np$ pairs contribute to the EMC effect, yielding an isospin-dependent effect in non-isoscalar nuclei.
In this case, the EMC effect should directly correlate with $a_2$~\cite{fomin2012}, which measures the relative contribution of high-momentum nucleons.

In the LD model, the EMC effect is taken to be isospin independent, with all short-distance $NN$ configurations contributing.
The EMC effect should then correlate to the number of short-distance pairs, independent of their isospin structure or their relative momentum. 
Ref.~\cite{fomin2012} extracted the number of SRCs, $R_{2N}$ by applying a correction to $a_2$ to account for the center-of-mass motion of the pair. As these SRCs are mainly $n$p pairs, we then scale $R_{2N}$ by the ratio of  $np$ pairs to total possible number of $NN$ pairs in the nucleus.
This analysis~\cite{Arrington:2012ax} found that the LD model gave a more direct proportionality between the EMC effect and SRC measurements, but the difference was not significant enough to claim that the isospin-independent explanation was significantly favored.

A recent work~\cite{Schmookler:2019nvf} performed a related analysis under the assumption that modification to the nuclear structure function can be written as a modification to the sum of nucleon structure functions that is proportional to the number of SRC pairs, $n_{SRC}^A (\Delta F_2^p+\Delta F_2^n)$.
Under these assumptions, this universal modification function can be derived from the measured EMC ratios and $a_2$:
\begin{equation}
F_{univ}^{HV}=\frac{(\sigma_A/\sigma_D)-(Z-N)\frac{F_2^p}{F_2^d}-N}{(A/2) a_2-N} .
\label{eq:hv}
\end{equation}
As in ref.~\cite{Schmookler:2019nvf}, isoscalar corrections are removed from published EMC ratios, according to the prescriptions used in the original data sets, and the $F_2^p/F_2^d$ term in Eq.~\ref{eq:hv} accounts for this effect in a unified fashion for all data sets. In this analysis, we take $F_2^p/F_2^d = F_2^p / (F_2^p+F_2^n)$ based on the $F_2^n/F_2^p$ ratio from ref.~\cite{Arrington:2008zh}. This is a good approximation for $x < 0.7$, the region we use to measure the size of the EMC effect.

\begin{figure}[ht]
\includegraphics[width=0.335\textwidth, angle=270]{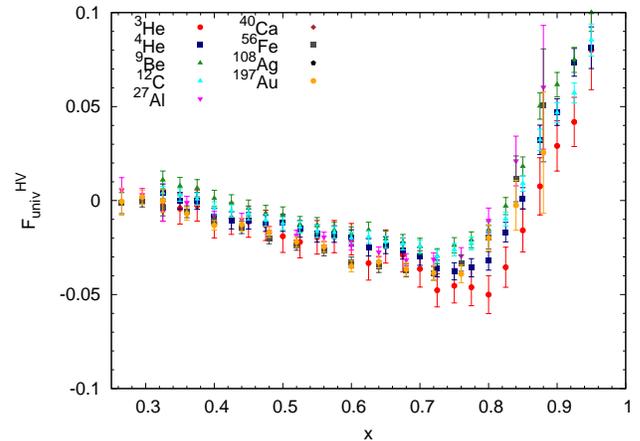}
\caption{Universal EMC effect function derived assuming np-dominance in the EMC effect. Note that for consistency with the analysis of Ref.~\cite{Schmookler:2019nvf}, the $^3$He ratio is shifted up by 3\% based on the analysis of Ref.~\cite{kulagin2010}.}
\label{fig:ratios_univHV}
\end{figure}

The universal function derived under these assumptions is shown for several nuclei in Fig.~\ref{fig:ratios_univHV}, yielding a fairly consistent underlying modification for all nuclei in the EMC effect region, $0.3 < x < 0.7$. The universality of this function was taken as evidence of an isospin-dependent EMC effect in Ref.~\cite{Schmookler:2019nvf}. While this provides a consistency check on the isospin-dependent assumption, the authors did not perform a similar check for an isospin-independent assumption to compare the two hypotheses. 

Following the approach taken in Ref.~\cite{Arrington:2012ax}, we extract universal functions to test both the HV and LD hypotheses. The HV hypothesis implies that the EMC effect is generated directly by the high-momentum contributions from SRC, yielding an excess EMC effect for protons in neutron-rich nuclei due to the $np$ dominance of SRCs, and is tested by the approach of Eq.~\ref{eq:hv}. We use a modified universal function to test the LD hypothesis, 
\begin{equation}
F_{univ}^{LD}=\frac{R_{EMC}-1}{R_{2N}\frac{A(A-1)}{2ZN}-1} .
\label{eq:ld}
\end{equation}
Note that, for consistency with the HV approach (Eq.~\ref{eq:hv}), isoscalar corrections were applied to all data sets in a consistent manner, i.e. we first remove the isoscalar corrections applied in the original publication, then applied a consistent isoscalar correction to all data:
\begin{equation}
C_{isospin}=\frac{Z F_2^p+N F_2^n}{A/2 (F_2^p+F_2^n)},
\end{equation}
where ${F_2^n}/{F_2^p}$ is taken from Ref.~\cite{Arrington:2008zh}. In eq.~\ref{eq:ld}, we start with $R_{2N}$, which represents the number of high-momentum pairs (dominantly $np$) and scale to the total number of potential $NN$ pairs with the factor $(A(A-1)/2)/(ZN)$.

\begin{figure}[htbp]
\center
\includegraphics[width=0.335\textwidth, angle=270]{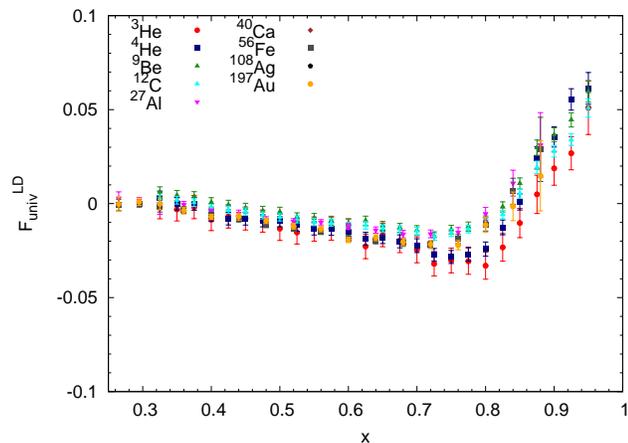}
\caption{Universal EMC effect function assuming a local density model with isospin independence.}
\label{fig:ratios_univLD}
\end{figure}

\begin{figure}[htbp]
\center
\includegraphics[width=0.325\textwidth, angle=270]{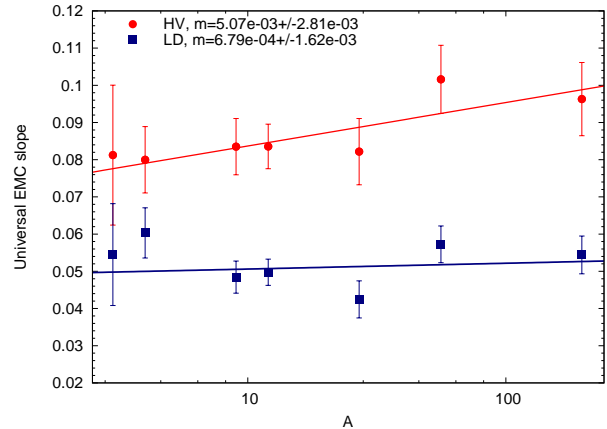}
\caption{Universal EMC slopes for the LD and HV analysis approaches as a function of $A$. Slopes are quoted from a fit to the universal function for $0.3<x<0.7$. The $A$ dependence for both universal functions is fit to the form $slope(A) = m \times ln(A) + b$.}
\label{fig:slopes_zoom}
\end{figure}

The LD-based universal function extracted according to Eq.~\ref{eq:ld} is shown in Fig.~\ref{fig:ratios_univLD}. As with the HV-based analysis, the data are in good agreement for $0.3 < x < 0.7$. To more carefully examine the `universality' of the HV and LD universal functions, we take the slope between $x=0.3$ and $x=0.7$ and plot this as a function of $A$ in Figure~\ref{fig:slopes_zoom}. A linear fit shows that the LD model yields a universal function that does not show a significant variation with $A$, while the HV model of Ref.~\cite{Schmookler:2019nvf} yields a slope that is nearly 2 standard deviations from zero. While this indication of an $A$ dependence under the HV assumption is relatively modest and cannot be taken as strongly favoring for the LD hypothesis, it cannot be said to favor the $np$-dominance model. 

Ref.~\cite{Schmookler:2019nvf} also examines the isospin dependence of the EMC effect in a second way. The authors modify the EMC effect slope by using the ratio of cross sections per proton (or neutron) rather than the conventional cross section ratio per nucleon.  The use of the slope as a measure of the EMC effect was proposed~\cite{Seely:2009gt} based on the observation that the EMC effect for $0.3 < x < 0.7$ had a universal $x$ dependence, with only the scale of the deviations from $R_{EMC}=1$ varying with the nucleus. In this definition, the slope is implicitly normalized by this `no EMC effect' expectation, $R_{EMC}=1$ when taking the cross section per nucleon.  The `no EMC effect' expectation will, of course, change if one uses a different cross section normalization, even though the relative modification of the nuclear pdf would not.

\begin{figure}[htbp]
\center
\includegraphics[width=0.4\textwidth, angle=270, trim=0 0 0 2in, clip]{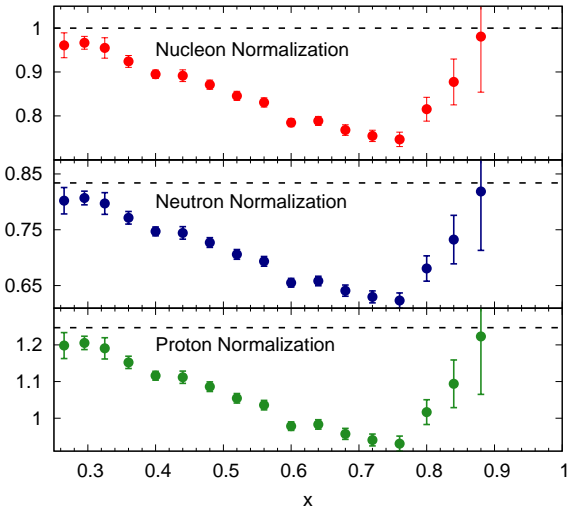}
\caption{EMC effect for gold, shown using several different normalizations. Top plot is the usual per nucleon ratio, $(\sigma_A/A)/(\sigma_D/2)$, middle bottom plot utilize per neutron and per proton normalizations, respectively: $(\sigma_A/N,Z)/(\sigma_D)$. The data are identical up to an overall normalization factor, and while the absolute slope scales with the normalization, the relative modification of the ratio relative to the `no EMC effect' values (indicated by dotted lines) are identical.}
\label{fig:emc_multiplot}
\end{figure}

To illustrate this, Figure~\ref{fig:emc_multiplot} shows the EMC ratios for gold, comparing the conventional per-nucleon cross section ratios to ones normalized per-proton and per-neutron. The ratios are identical up to a scaling factor. This scaling factor changes the absolute slope in the EMC effect region, but this does not reflect a change in the relative nuclear pdf modification. Making explicit the normalization to the `no EMC effect' ratio, one recovers identical modifications in each case.

While there are many reasons to expect an isospin-dependent EMC effect in non-isoscalar nuclei~\cite{cloet09, Arrington:2015wja}, there are presently no data that isolate the proton and neutron (or up- and down-quark) contributions. Examining the EMC-SRC correlation under different assumptions about their microscopic connection is one way to attempt to compare the HV ($np$-dominance) and LD (isospin-independent) pictures of the underlying physics.  An initial in-depth examination~\cite{Arrington:2012ax} of the correlation suggested that the LD model yielded a better correlation, but the differences were not large enough to argue that it was strongly favored over the HV model. Our new analysis, following Ref.~\cite{Schmookler:2019nvf} but examining both the LD (isospin independent) and HV (isospin dependent) hypotheses again favors the LD model, in that the data are consistent with a truly universal function, while the HV model is best fit with a small $A$ dependence.  Additional data are needed to study the isospin dependence of the EMC effect.  Clear knowledge of this dependence would impact other measurements utilizing DIS from nuclei~\cite{cloet09,Cloet:2012td}, as well as prove insight into the origin of the EMC effect and its connection to the SRCs.

Upcoming approved Jefferson Lab experiments will better map out the EMC-SRC correlation, increasing the sensitivities of the tests presented here. Measurements of the EMC effect~\cite{E1210008} and SRCs~\cite{E1206105} in both light ($A \le 12$) nuclei and medium-to-heavy ($A \ge 40$) nuclei, will provide a broader data set with which to examine the EMC-SRC correlation. In addition, these experiments will include nuclei that vary the neutron-to-proton ratio for nuclei with similar masses, allowing a cleaner separation of the $A$ dependence and $N/Z$ dependence of both SRCs and the EMC effect.

A more direct test is a pdf comparison of mirror nuclei, e.g. using recently taken data on $^3$H and $^3$He~\cite{e1210103}. Here, the isospin structure of the target appears in two ways: as a difference between the free proton and neutron structure functions and through the potential difference in their modification in the nucleus. Detailed studies suggest that this comparison is more sensitive to the ratio $F_2^n/F_2^p$~\cite{afnan03}. In the future, a precise and model-independent understanding of $F_2^n/F_2^p$ for free nucleons would allow direct searches for an isospin-dependent EMC effect.

Other paths to probe the flavor dependence of the EMC effect include pion-induced Drell-Yan scattering~\cite{Dutta:2010pg} and flavor tagged measurements of the EMC effect~\cite{PR1209004,Chang:2011ra,E1217012}. Perhaps the cleanest way to isolate the isospin dependence of the EMC effect is through measurements of parity-violating deeply inelastic scattering (PVDIS) from non-isoscalar nuclei. For this measurement, the parity-violating asymmetry is well understood except for the potential modification of the up- and down-quark distributions in a non-isoscalar nuclei~\cite{Cloet:2012td, cloet19}. Such a measurements would be possible, e.g. on $^{48}$Ca, using the SoLID spectrometer proposed for PVDIS measurements on the deuteron and proton~\cite{E1210007, PR1216006}.

In conclusion, while interesting and suggestive data exist, the exact nature of the connection between the EMC effect and SRCs has not been illuminated, nor has the isospin dependence or lack thereof for the former been established.  The upcoming suite of EMC effect measurements at Jefferson Lab is designed elucidate the nature of the EMC effect and its potential connection to the short-range structure of the nucleus by examining a range of sensitive observables~\cite{cloet19}.

This work was supported by the U.S. Department of Energy, Office of Science, Office
of Nuclear Physics, under contracts DC-AC02-06CH11357 and DE-SC0013615.

\bibliography{emc_univ}

\end{document}